\documentclass[12pt]{iopart}
\usepackage{iopams}
\usepackage{graphicx}   
\usepackage{bm}

\def\brCP{\mathbf r_{CP}}
\def\btau{\boldsymbol\tau}
\def\hbs{\hat{\mathbf s}}
\def\hbt{\hat{\boldsymbol\theta}}
\def\hbx{\hat{\mathbf x}}
\def\hby{\hat{\mathbf y}}
\def\hbz{\hat{\mathbf z}}

\begin{document}

\title{Rolling of asymmetric disks on an inclined plane}

\author{Ben Yu-Kuang Hu}
\address{Department of Physics, University of Akron, Akron,
OH~44325-4001, U.S.A.} \ead{byhu@akron.edu}

\begin{abstract}
In a recent papers, Turner and Turner (2010 {\em Am.~J.~Phys.}
{\bf 78} 905-7) and Jensen (2011 {\em Eur.~J.~Phys.} {\bf 32}
389-397) analysed the motion of asymmetric rolling rigid bodies on
a horizontal plane.  These papers addressed the common
misconception that the instantaneous point of contact of the
rolling body with the plane can be used to evaluate the angular
momentum $\mathbf L$ and the torque $\boldsymbol\tau$ in the
equation of motion $d\mathbf L/dt = \boldsymbol\tau$.  To obtain
the correct equation of motion, the ``phantom torque" or various
rules that depend on the motion of the point about which $\mathbf
L$ and $\boldsymbol\tau$ are evaluated were discussed. In this
paper, I consider asymmetric disks rolling down an inclined plane
and describe the most basic way of obtaining the correct equation
of motion; that is, to choose the point about which $\mathbf L$
and $\boldsymbol\tau$ are evaluated that is stationary in an
inertial frame.
\end{abstract}

\date{\today}

\maketitle

In a recent papers, Turner and Turner \cite{turner} and Jensen
\cite{jensen} discussed the dynamics of a rolling body which is
not azimuthally symmetric on a horizontal plane and the standard
technique described in introductory physics textbooks for
obtaining the equation of motion, which is to use $d\mathbf L/dt =
\boldsymbol\tau$, where $\mathbf L$ is the angular momentum and
$\boldsymbol\tau$ is the torque with respect to the contact point
of the rolling body with the plane, gives incorrect results. They
derived the correct equation of motion, either by introducing an
additional term which they refer to as a phantom torque
\cite{turner} or by using rules \cite{jensen} that depend on the
motion of the point about which $\mathbf L$ and $\boldsymbol\tau$
are evaluated.

The phantom torque in Ref.~\cite{turner} was needed because the
point about which the torques and the angular momenta were
evaluated, the contact point between the hoop and the plane, was
assumed to be attached to the semicircular hoop (see Fig.~2 in
Ref.~\cite{turner}), and hence it accelerates as the hoop rolls.
The equation of motion $d\mathbf L/dt = \btau$, is in general
invalid when this point is accelerating respect to an inertial
frame, since it is based on Newton's second law which is valid
only in an inertial frame. However, in the same way that the
validity of Newton's second law can be restored in an accelerating
(non-rotating) frame of reference by inclusion of a fictitious
inertial force term, so can that of $d\mathbf L/dt =
\boldsymbol\tau$ be with a phantom torque term as described in
Ref.~\cite{turner}.  Indeed, as mentioned in Ref.~\cite{jensen},
the phantom torque Eq.~(14) in Ref.~\cite{turner} can be obtained
from the fictitious inertial force of an accelerated frame.
Namely, in the frame of reference of a point $P$, the mass
elements on an object, $dm$, experience a fictitious force
$d\mathbf F_{\mathrm{fict}} = -\mathbf a_{P}\,dm$, where $\mathbf
a_{P}$, is the acceleration of $P$ with respect to an inertial
frame \cite{accelframe}. This results in a phantom torque
$\boldsymbol\tau_{\mathrm{ph}} = \int (\mathbf r - \mathbf r_{P})
\times d\mathbf F_{\mathrm{fict}} = \int (\mathbf r-\mathbf
r_{P})\times (-\mathbf a_{P}\,dm) = -M \brCP \times \mathbf
a_{P}$, where $M$ is the total mass of the object and $\mathbf
r_{CP}$ is the position of its center of mass with respect to $P$,
since $\int (\mathbf r-\mathbf r_{P})\ dm = M \mathbf r_{CP}$.

Ref.~\cite{jensen} introduces two different rules to take into
account the motion of the point of reference about which $\mathbf
L$ and $\boldsymbol\tau$ are evaluated.  The second rules, given
in Eq.~(10) of Ref.~\cite{jensen} is equivalent to the phantom
torque technique of Ref.~\cite{turner}.

This leads naturally to the question, why not simply solve the
problem using the most basic method; that is, using the equation
of motion $d\mathbf L/dt = \boldsymbol\tau$ where $\mathbf L$ and
$\boldsymbol\tau$ are evaluated about a point $P$ that is
stationary with respect to an inertial frame? I consider this
method in this note, for the more general case than in
Refs.~\cite{turner} and \cite{jensen}, that of a planar disk of
radius $R$ and mass $M$ rolling down an inclined plane, which is
tilted at an angle $\gamma$ relative to horizontal.  The disk does
not necessarily have uniform density, and has a center of mass
that is a distance $\beta R$ from the geometric center of the disk
(where $0 \leq \beta < 1$).  The friction of the inclined plane is
sufficient to prevent the disk from slipping.  Let $\theta$ be the
angle of the angle of rotation of the cylinder in the
counter-clockwise direction, and $\theta = 0$ correspond to the
case where the center of mass, $C$, the point of contact of the
cylinder with the inclined plane, $P$, and the geometric center of
the cylinder, $O$, are co-linear. Define $\hbx$, $\hby$ and $\hbz$
to be the unit vectors in the directions of motion of the disk,
perpendicular to the inclined plane, and perpendicular to the
surface of the disk, respectively.  Let the unit vector $\hbs$ to
be in the direction from $O$ to $C$, and $\hbt$ to be the unit
vector perpendicular to $\hbs$, as shown in the Fig.~1. In this
note, I use double subscripts in variables to denote ``with
respect to". For example, $\mathbf r_{CP}$ is the position of $C$
with respect to $P$.

The point $P$ is assumed to be {\em attached to the plane on which
the disk is rolling}, and hence is stationary with respect to an
inertial frame.  The angular momentum of the object with respect
to a point $P$ can be separated \cite{taylor} into an ``orbital"
part $\mathbf L_{\mathrm{orb}}$, the angular momentum of the
center of mass with respect to $P$ and a ``spin" part $\mathbf
L_{\mathrm{spin}}$, the angular momentum relative to the center of
mass of the object, \numparts
\begin{eqnarray}
\mathbf L &= \mathbf L_{\mathrm{orb}} + \mathbf L_{\mathrm{spin}}
\label{eq:L}\\\mathbf L_{\mathrm{orb}} &= M \brCP \times \mathbf
v_{CP}\\ \mathbf L_{\mathrm{spin}} &= {\mathbb{I}}_{C} \cdot
\boldsymbol\omega,
\end{eqnarray}
\endnumparts
where $\mathbf v_{CP}$ is the velocity of the center of mass with
respect to $P$, ${\mathbb{I}}_{C}$ is the moment of inertia tensor
relative to the center of mass and $\boldsymbol\omega$ is the
angular velocity vector.  Using Eqs.~(1) in the equation of motion
$d\mathbf L/dt = \boldsymbol\tau$, for the case considered here
\cite{no_euler_term}, gives
\begin{equation}
{\mathbb I}_{C}\cdot \dot{\boldsymbol \omega} + M \mathbf r_{CP}
\times \mathbf a_{C} = \boldsymbol\tau\label{eq:2}
\end{equation}
where $\mathbf a_{C}$ is the acceleration of the center of mass
relative to an inertial frame.

The first term on the left hand side of Eq.~(\ref{eq:2}) is simply
$I_{C}\ddot\theta\hbz$, where $I_C$ is the moment of inertia
through $C$ in the $z$-direction. To determine the second term, we
need to find $\mathbf a_C$ as a function of $\theta$. The position
of the center of mass $C$ relative to the $\theta = 0$ point of
the contact of the disk and the inclined plane is $R(-\theta
\hat{\mathbf x} + \hat{\mathbf y} + \beta\hat{\mathbf s})$; {\em
i.e.}, the position of $O$ relative to the point of contact for
$\theta = 0$ plus the position of $C$ relative to $O$. Taking the
second derivative with respect to time, and using the result for
the angular and centripetal accelerations for circular motion
\cite{circularmotion}, gives $\mathbf a_C = R(-\ddot\theta \hbx +
\beta\ddot\theta\hbt -\beta\dot\theta^2\hbs).$ This, together with
$\brCP = R(\hat{\mathbf y} + \beta\hat{\mathbf s})$, gives
\begin{eqnarray}
M \brCP \times \mathbf a_C &= MR^2 (\hat{\mathbf y} + \beta\hbs)
\times (-\ddot\theta \hat{\mathbf x} + \beta\ddot\theta\hbt -
\beta\dot\theta^2\hbs )\nonumber\\
&= MR^2 \left[ (1 + \beta^2 - 2\beta\cos\theta) \ddot\theta +
 \beta\sin\theta \dot\theta^2 \right]\hbz ,
\end{eqnarray}
where $\hbs = \sin\theta\,\hbx - \cos\theta\,\hby$ and $\hbt =
\cos\theta\,\hbx + \sin\theta\,\hby$ is used to evaluate the cross
products.
The direction of the weight force, in the case of an inclined
plane that is tilted with an angle of $\gamma$, is $\mathbf
F_{\mathrm{wt}} = -Mg (\sin\gamma \hbx +\cos\gamma \hby)$ and
therefore the torque with respect to $P$ is
\begin{equation}
\btau = \brCP \times \mathbf F_{\mathrm{wt}} = MgR
\left[\sin\gamma - \beta \sin(\theta + \gamma)\right]\hbz.
\end{equation}
(The normal and frictional force of the inclined plane on the disk
act through point $P$, so do not contribute to the torque with
respect to $P$.)   Thus, Eq.~(\ref{eq:2}) can be written as
\begin{equation}
I_{P}\ddot\theta + MR^2\beta \sin\theta\,\dot\theta^2 = MgR
\Bigl[\sin\gamma - \beta\sin(\theta + \gamma)\Bigr],\label{eq:5}
\end{equation}
where $I_{P} = I_C + M(1+\beta^2 - 2\beta \cos\theta)R^2$ is the
moment of inertia about point $P$ by the parallel axis theorem,
since $(1 + \beta^2 - 2\beta\cos\theta)R^2 = \vert\brCP\vert^2$.
For the case of a uniform semicircular hoop on a level plane,
where $\gamma = 0$, $\beta = 2/\pi$  and $I_P = 2MR^2(1 -
\frac2\pi \cos\theta)$, Eq.~(\ref{eq:5}) reproduces the results
obtained in Refs.~\cite{turner} and \cite{jensen}.

The approach typically adopted by introductory texts is to use
$d\mathbf L/dt = \boldsymbol\tau$ where $\mathbf L$ and
$\boldsymbol\tau$ are evaluated with respect to $P$.  However,
typically the approach incorrectly mixes two methods -- the point
$P$ is assumed to be stationary in an inertial frame, but in
evaluating $d\mathbf L/dt$, the point $P$ is assumed to be
attached to the disk, giving in general the incorrect equation of
motion $I_P\ddot\theta = \tau$.  It misses the term
$MR^2\beta\sin\theta\,\dot\theta^2$, which can be generated by
including a phantom torque term that arises from the acceleration
of point $P$.  When the point $P$ is compelled to be stationary
with respect to an inertial frame, the phantom torque vanishes.
Nevertheless, the $MR^2\beta\sin\theta\,\dot\theta^2$ still
appears, coming instead from the centripetal acceleration of $C$
of the disk relative to $O$.  This term vanishes when $O$ and $C$
are coincident, as in the standard case discussed in introductory
physics textbooks, which is why this error has escaped detection.
In a sense, the error is a subtle case of the standard
freshman-level mistake of confusing zero velocity with zero
acceleration.  Just because the point of contact between the disk
and the plane is instantaneously stationary does not mean that it
can be used as a point of reference for the equation of motion
$d\mathbf L/dt = \boldsymbol\tau$.  The point of reference must
either have a zero acceleration with respect to an inertial frame,
as in this note, or if it has non-zero acceleration, as discussed
in in Refs.~\cite{turner} and \cite{jensen}, additional terms must
be introduced.



\noappendix
\newpage
\section*{Figures}
\begin{figure}[h]
\begin{center}
\includegraphics{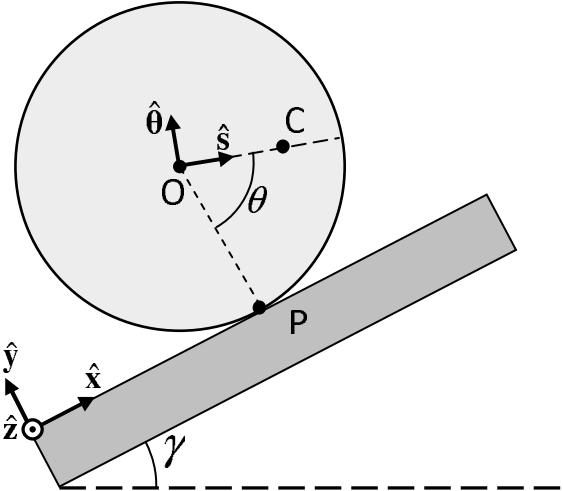}
\end{center}
\caption{Rolling non-symmetric disk on a plane inclined at an
angle $\gamma$ to horizontal.
 $O$ is the geometric center, $C$ is the center of mass, $P$ is the point of contact of the disk with the inclined plane, and $\theta$ is the angle
 of the line through $O$ and $C$ with respect to perpendicular to the plane.  Unit vector $\hbs$ points from $O$ to $C$, and $\hbt = d\hbs/d\theta$.}\label{fig:averageG}
\end{figure}

\end{document}